\documentstyle{article}
\begin{document}
\begin{center}
{\Large\bf Massive Electrodynamics and Magnetic Monopoles.
}\\[1cm]
{\bf
Mark Israelit$^{\dagger, }$\footnote{e-mail:
 Marc.Israelit@uni-konstanz.de , and:   
israelit@physics.technion.ac.il , (permanent e-address)}
}\\[0.5cm]
$^\dagger$ Department of Physics, University of Konstanz,
PF 5560 M678, D-78434 Konstanz, Germany.
On leave {F}rom: Department of Physics, University of Haifa,
Oranim, Tivon 36006, Israel.\\
\medskip
\end{center}
\begin{abstract}
\noindent
Including torsion in  the geometric framework of the Weyl-Dirac
theory we  build up an action integral, and obtain {F}rom it a
gauge covariant (in the Weyl sense) general relativistic massive    
electrodynamics. Photons having an arbitrary mass, electric,
and magnetic currents (Dirac's monopole)  coexist within
this theory. Assuming that the space-time is torsionless,
taking the photons mass zero, and turning to the Einstein
gauge we obtain  Maxwell's electrodynamics.
\end{abstract}
\vspace{0.5cm}
\section{Introduction}
Ever after Dirac put forward the idea of magnetic poles
\cite{Dirac1931},  there has been continued interest and
discussion on this subject. It is  not surprising that
this interest increased soon after Dirac \cite{Dirac1948}
succeeded in setting up a more general theory, which led to
a dynamical justification of his famous quantization
condition. {F}rom an interesting review article by
Blagojevi\'{c}  and Senjanovi\'{c} \cite{Blagojevic},
which covers over 150 works  dealing with monopoles
during  six decades,  one can see that   theoretically
oriented papers are dealing mostly  with quantum aspects
of the phenomenon, while the problem of constructing a
satisfactory classical general relativistic  framework
was ignored in almost  all works .

However, the problem must be considered also {F}rom the
classical standpoint. First of all in order to
quantize the fields it is necessary to have an action
integral that provides us with a non-contradictory
classical theory. Secondly, there are of course
classical aspects of the monopole phenomenon.
An example might be  the  explanation of  magnetic
fields of celestial bodies.
Finally, if a magnetic charge (monopole) really exists,
then  Maxwell's  electrodynamics that suffers {F}rom an
asymmetry, regarding to electric, and   magnetic currents,
must be replaced by a generalized theory having a
non-vanishing dual field tensor. Very interesting
attempts in this direction  were made by  Hammond
\cite{Hammond}, who tried to describe electric and
magnetic fields by torsion.

Another interesting problem is the photons mass.
The massless photon became a tacit axiom of physics
own to the success of quantum electrodynamics in
predicting experiments with  enormously high exactness.
But the same results would be obtained with
photons having mass  $m_{\gamma}< 10^{-48}\hbox{g}$
(cf. \cite{Goldhaber}). Moreover, as recently noted
Ignatiev and Joshi  \cite{Ignatiev}, in some respects
massive QED is simpler theoretically than the standard
theory, and {F}rom the experimental point of view
massive QED is as perfect as standard QED.

The purpose of this paper is to present a generalized
eletrodynamics allowing both, electric and magnetic
currents, and also massive photons. The theory is
based on a generalization of  the Weyl-Dirac geometry
\cite{Weyl1919},  \cite{Dirac1973}. The  equations are
derived {F}rom a geometrically justified  action integral,
and in the limiting case one  has the ordinary
Einstein-Maxwell theory.
\section{The Geometric Basis}
In order to develop an appropriate geometric
basis we shall generalize the geometry of the
Weyl-Dirac theory \cite{Weyl1919}, \cite{Dirac1973}
by including  a contorsion term  in the connection.
We assume that there is a symmetric  metric
tensor $g_{\mu\nu}$, and an assymetric connection
$\Gamma^\lambda_{\,\mu\,\nu}$ in  each point of
the manifold . The connection  will be split into
three parts, the Christoffel symbol, the non-metricity,
and the contorsion. With the non-metricity having the
Weylian form one can  write (cf. \cite{Schouten})
\begin{equation} \label{1}
\Gamma^\lambda_{\,\mu\,\nu}=
\left\{^\lambda_{\mu\,\nu}\right\}+g_{\mu\nu} w^\lambda
-   \delta^\lambda_\mu w_\nu -\delta^\lambda_\nu w_\mu
  +C^\lambda_{\ \mu\,\nu}  ,
\end{equation}
where $w_\mu$  is  the Weyl connection vector,
and with the contorsion given  in terms of the
torsion tensor as follows
\begin{equation} \label{2}
C^\lambda_{\,\mu\nu}=
\Gamma^\lambda_{\,\left[\, \mu\nu\right]}-g^{\lambda\beta}
g_{\sigma\mu}\Gamma^\sigma_{\,\left[\beta\nu\right]}-g^{\lambda\beta}g_{\sigma\nu}\Gamma^\sigma_{\,\left[\beta\mu\right]},
\end{equation}
Actually one has now a Weyl geometry \cite{Weyl1919}
with torsion, so that both, the direction and the
length of a vector may change in the process of
parallel displacement. For a vector $B^\mu$,
having the length $B$, these changes are given by
\begin{equation}\label{3}
dB^\mu= -B^\sigma\Gamma^\mu_{\,\sigma\nu}dx^\nu,
\end{equation}
and
\begin{equation} \label{4}
dB=B w_\nu dx^\nu .
\end{equation}
It follows that, after a parallel displacement
around a parallelogram formed by $dx^\mu$ and
$\delta x^\nu$, the length changes by
\begin{equation}\label{5}
\Delta B =-BW_{\mu \nu}dx^\mu \delta x^\nu ,
\end{equation}
with
\begin{equation}\label{6}
W_{\mu \nu}=  w_{\mu , \nu}-w_{\nu , \mu}.
\end{equation}

This nonintegrability of length leads to an arbitrary
standard of length, or gauge, at each point and hence to
local gauge transformations
\begin{equation}\label{7}
B\rightarrow\overline B=e^\lambda B\ ; \qquad g_{\mu\nu}
\rightarrow\overline g_{\mu\nu}=e^{2\lambda}g_{\mu\nu} ; \qquad  
w_\mu\rightarrow\overline w_\mu=w_\mu+\lambda_{, \mu }.
\end{equation}
with  $\lambda(x^\mu)$  being an arbitrary function.
The third relation in (\ref{7}) is the familiar transformation
that one has in the Maxwell theory  for the electromagnetic
potential vector, and it led Weyl to identify $w_{\mu}$ with
this vector, so that one has a connection between
electromagnetism  and geometry. In our generalization
of the Weyl geometry we will  assume  that the torsion
tensor is gauge invariant, so that in addition to (\ref{7}) we have
\begin{equation}\label{A}
\Gamma^\lambda_{\,\left[\,\mu\nu\right]}  \rightarrow
\overline{\Gamma}^\lambda_{\,\left[\,\mu\nu\right]} =
\Gamma^\lambda_{\,\left[\,\mu\nu\right]} .
\end{equation}
    Considering a parallel displacement around an infinitesimal
parallelogram, one can derive the curvature tensor
\begin{equation}\label{8}
K^\lambda_{\mu\nu\sigma}=- \Gamma^\lambda_{\mu\nu,\sigma}
 +\Gamma^\lambda_{\mu\sigma,\nu}  -
\Gamma^\alpha_{\mu\nu}  \Gamma^\lambda_{\alpha\sigma} +
\Gamma^\alpha_{\mu\sigma}  \Gamma^\lambda_{\alpha\nu}.
\end{equation}
The two curvature tensors (\ref{6}), and (\ref{8}) may
be used to build up an action integral. In order to get
an action integral, which is invariant under both, the coordinate  
transformations, and the gauge transformations (\ref{7}),
and which agrees with the Einstein general relativity,
Dirac \cite{Dirac1973} introduced a scalar
field $\beta(x^\mu)$. It is assumed that under a gauge
transformation (\ref{7}) the  gauging
function $\beta(x^\mu)$ changes according to
\begin{equation}\label{9}
\beta\rightarrow\overline\beta = e^{-\lambda}\beta.
\end{equation}
For the torsionless case Dirac built up a geometrically
based action. He made use of the curvature scalar  
$K=g^{\mu\nu}K^\lambda_{\mu\nu\lambda}$ formed {F}rom (\ref{8}),
of a term $W_{\mu\nu} W^{\mu\nu}$  constructed {F}rom the
Weylian length curvature tensor (\ref{6}), as well as of
the vector $w_{\mu}$ and the scalar $\beta$. This action
has the following form \cite{Dirac1973}
\begin{equation}\label{10}
I_{Dirac} =\int \left[W^{\mu\nu}W_{\mu\nu}- \beta^2  
K+k(\beta_{,\underline\mu}+\beta w^\mu) (\beta_{,\mu}
+\beta w_\mu)+ 2\Lambda \beta^4 +L_{matter}\right](-g)^{1/2}\,dx
\end{equation}
where $\Lambda$ is the cosmological constant, $k$ an
arbitrary parameter, $L_{matter}$ is the Lagrangian
density of matter, and an underlined index is to be
raised with $g^{\mu\nu}$. In order to get {F}rom (\ref{10})
the Maxwell field equations Dirac took $k=6$. Later
Rosen \cite{Rosen1982} gave a detailed discussion
on the Weyl-Dirac theory. He  pointed out that
for $k\neq 6 $, one  obtains a Proca equation \cite{Proca1936}
instead of Maxwell's one.For $k-6<0$ {F}rom the standpoint of
quantum mechanics the Proca field may be interpreted as an
ensemble of bosons, particles of spin 1 and finite mass.
These massive vector bosons , named weylons, were used for  
obtaining  non-baryonic dark matter in the universe  
\cite{Isr.Rosen1992}, \cite{Isr.Rosen1994} .

In the present work it will be assumed that electromagnetism
may be described by the Proca equations, so that the particles
will be regarded rather as massive photons. Moreover, we
incorporate torsion into the space, so that the
connection (\ref{2}), as well as the curvature (\ref{8})
contain torsion. For a moment let us consider  the second
derivative of a  given vector $B_\mu$, with respect to the  
connection $\Gamma^\lambda_{\mu\nu}$ defined in (\ref{1}). (cf.   
\cite{Schouten}):
\begin{equation}\label{11}
B_{\mu:\nu:\sigma}- B_{\mu:\sigma:\nu} =B_\lambda
K^\lambda_{\mu\nu\sigma}
-2B_{\mu:\alpha} \Gamma^\alpha_{\,\left[\nu\sigma\right]}.
\end{equation}
In the last term of (\ref{11}) appear additional geometric
properties, that are not expressed neither in (\ref{6}), nor in  
(\ref{8}).
In order to include them in the Lagrangian one has to
replace  $B_\mu$ by a fundamental geometric quantity that is
already present in the framework. For this purpose let us take the  
tensor   $W_{\mu\nu}$ (It may be   recalled here that   
$g_{\mu\nu:\lambda} =-2g_{\mu\nu}w_\lambda$, so that $W_{\mu\nu}$  
may be regarded as the
second derivative of $g_{\mu\nu}$). By this choice  the  torsional
curvature term takes on the form
\begin{equation}\label{12}
aW_{\mu\lambda:\alpha} \Gamma^\alpha_{\,\left[\nu\sigma\right]}
\end{equation}
with  $a$  being an abitrary constant.

In the next section we will make use of (\ref{6}), (\ref{8}),
and (\ref{12}) for building up an action .

\section{ Variational  Principle }

Let us start from Dirac's action (\ref{10}). It is convenient to
express the curvature scalar $K$ explicitly in terms of the  
Christoffelsymbol $\left\{^\lambda_{\mu\,\nu}\right\}$, the Weyl  
connection vector $w_\mu$, and the torsion tensor  
$\Gamma^\lambda_{\,\left[\mu\nu\right]}$ according to (\ref{1}), and  
(\ref{2}). We will include also the scalar  $aW_{\mu\nu;\alpha}  
\Gamma^\alpha_{\,\left[\lambda\sigma\right]}  
g^{\mu\lambda}g^{\nu\sigma}$, stemming from (\ref{1}\ref{2}), but  
with  $\Gamma$-differentiation (:) replaced  by the Riemannian one  
(;)
(formed with $\left\{^\lambda_{\mu\,\nu}\right\}$) . Thus we have
the following action integral:
\begin{eqnarray}\label{13}
I =\int \biggl(W^{\mu\nu} W_{\mu\nu} -\beta^2 R+\beta^2(k-6)w_\mu  
w^\mu + 2(k-6)\beta w^\mu \beta_{,\mu}+ k\beta_{,\mu}  
\beta_{,\underline\mu}  +8  
\beta\Gamma^\alpha_{\,\left[\lambda\alpha\right]}
\beta_{,\underline \lambda}
\\
\nonumber
+\beta^2(2\Gamma^\alpha_{\,\left[\mu\lambda\right]}
\Gamma^ \lambda_{\,\left[{\underline\mu} \alpha\right]} -   
\Gamma^\alpha_{\,\left[\sigma\alpha\right]}
\Gamma^\omega_{\,\left[{\underline\sigma}\omega\right]}   +          
                     \Gamma^\alpha_{\,\left[\mu\lambda\right]}
\Gamma^\omega_{\,\left[\underline\mu\underline\lambda\right]}
g_{\alpha\omega} + 8\Gamma^\alpha_{\,\left[\sigma\alpha\right]}
w^\sigma)
\\
\nonumber
+aW_{\mu\nu;\alpha}
\Gamma^\alpha_{\,\left[\underline\mu\underline\nu\right]}
+2\Lambda \beta^4 +L_{matter}\biggr)(-g)^{1/2}\,dx
\end{eqnarray}
with $R$ being the Riemannian curvature scalar.

One can readily prove ({F}rom the field equations) that $a=4$.
Below  this value will be  taken. Varying in (\ref{13}) $w_\mu$
one obtains the following field equation
\begin{equation}\label{14}
\biggl(W^{\mu\nu} -
 2\Gamma^\alpha_{\,\left[\underline{\mu}    
\underline{\nu}\right];\alpha}\biggr)_{;\nu} =  
(1/2)\beta^2(k-6)W^\mu +   
2\beta^2\Gamma^\alpha_{\,\left[\underline{\mu}\alpha\right]} +4\pi  
J^\mu
\quad ,
\end{equation}
where $W_\mu$ stands for the gauge-invariant Weyl vector  (cf. (\ref{7}))
\begin{equation}\label{15}
W_\mu=w_\mu+(\ln\beta)_{,\,\mu}\quad ,
\end{equation}
and
\begin{equation}\label{16}
16\pi J^{\mu}=\frac{\delta L_{matter}}{\delta w_{\mu}}\quad .
\end{equation}
Considering the variation of (\ref{13}) with respect to   
$\Gamma^\lambda_{\,\left[\mu\nu\right]}$ we obtain a second field  
equation :
\begin{eqnarray}\label{17}
W^{\mu\nu}_{;\lambda}= \beta^2(\delta^{\mu}_{\lambda}
 W^{\nu}-\delta^{\nu}_{\lambda} W^{\mu})+  
\beta^2(\delta^{\nu}_{\lambda}              g^{\mu\sigma}-  
\delta^{\mu}_{\lambda} g^{\nu\sigma})   
\Gamma^\alpha_{\,\left[\sigma\alpha\right]}
\\
\nonumber
 +(1/2)\beta^2(g^{\sigma\nu} \delta^\mu_\omega\delta^\rho_\lambda   
- g_{\lambda\omega}g^{\sigma\mu}g^{\rho\nu}-  
g^{\sigma\mu}\delta^\nu_\omega \delta^\rho_\lambda)  
\Gamma^\omega_{\,\left[\sigma\rho\right]} -                 
4\pi\Omega^{\,\left[\mu\nu\right]}_{\lambda} \quad,
\end{eqnarray}
where
\begin{equation}\label{18}
16\pi\Omega^{\,\left[\mu\nu\right]}_{\lambda} =\frac{\delta  
L_{matter}}{\delta\Gamma^\lambda_{\,\left[\mu\nu\right]}} \quad .
\end{equation}

The form of eq. (\ref{14}) justifies introducing the field
\begin{equation}\label{19}
\Phi_{\mu\nu}=W_{\mu\nu}-2\Gamma^\alpha_{\,\left[\mu\nu\right];
\alpha}\equiv
W_{\mu;\nu}-W_{\nu;\mu}-2\Gamma^\alpha_{\,\left[\mu\nu\right];\alpha}\quad.
\end{equation}

If one varies in (\ref{13}) the metric tensor $g^{\mu\nu}$, one  
obtains the following equation for the gravitational field:
\begin{eqnarray}\label{20}
\beta^2 G^{\mu\nu}&=&-8\pi T^{\mu\nu}
-8\pi(\widetilde{M}^{\mu\nu} -\overline{M}^{\mu\nu})                 
                                         + 2\beta(g^{\mu\nu}  
\beta_{;\alpha ;\underline{\alpha}} -
\beta_{;\underline{\mu}; \underline{\nu}})
+ 4\beta_{,\underline{\mu}} \beta_{,\underline{\nu}}-
 g^{\mu\nu}\beta_{,\underline{\alpha}} \beta_{,\alpha}
\\
\nonumber
&+&(k-6)\beta^2(W^\mu W^\nu- 1/2g^{\mu\nu}W^\sigma W_\sigma)
+4\beta^2 \Gamma^\alpha_{\,\left[\sigma
\alpha\right]}\, (g^{\sigma\nu}W^\mu
+ g^{\sigma\mu}W^\nu - g^{\mu\nu}W^\sigma)
\\
\nonumber
&+& \beta^2 \Gamma^\alpha_{\,\left[\sigma\tau \right]}
\Gamma^\omega_{\,\left [\lambda\rho\right]} f(g ;\delta)
-g^{\mu\nu} \beta^4 \Lambda  \quad ,
\end{eqnarray}
with
\begin{eqnarray}
f(g;\delta)= 2\delta^{\tau}_{\alpha}\delta^{\rho}_{\omega}
(g^{\lambda\sigma}g^{\mu\nu} -g^{\lambda\mu}g^{\sigma\nu} -
g^{\lambda\nu}g^{\sigma\mu})+ \delta^{\tau}_{\omega}
\delta^{\rho}_{\alpha}
(g^{\lambda\mu}g^{\sigma\nu} +g^{\lambda\nu}g^{\sigma\mu} -
g^{\lambda\sigma}g^{\mu\nu} )
\nonumber
\\
+g^{\tau\rho}(2g^{\lambda\mu} g^{\sigma\nu} g_{\alpha\omega}  
-\delta^{\mu}_{\omega}
\delta^{\nu}_{\alpha}g^{\lambda\sigma} -\frac{1}{2}g^{\mu\nu}
 g^{\lambda\sigma}g_{\alpha\omega}),
\nonumber
\end{eqnarray}
and where $8\pi T^{\mu\nu}=\delta L_{matter}/\delta g_{\mu\nu}$.
The modified energy density tensors of the field are defined as follows
\begin{equation}\label{21}
4\pi\widetilde{M}^{\mu\nu} =(1/4)g^{\mu\nu}\Phi^{\alpha\beta}  
\Phi_{\alpha\beta} - \Phi^{\mu\alpha}\Phi^\nu_{\, \alpha} \quad ,
\end{equation}
and
\begin{equation}\label{22}
4\pi\overline{M}^{\mu\nu} =(1/4)g^{\mu\nu}(\Phi^{\alpha\beta}  
-W^{\alpha\beta})
(\Phi_{\alpha\beta}-W_{\alpha\beta} ) - (\Phi^{\mu\alpha}-W^{\mu\alpha})
(\Phi^\nu_{\,\alpha}-W^\nu_{\,\alpha}) \quad .
\end{equation}
Contracting eq. (\ref{20}) we get:
\begin{eqnarray}\label{23}
&R& -(8\pi/\beta^2) T+(6/\beta) \beta_{;\alpha;\underline{\alpha}}  
-(k-6) W^\sigma
W_\sigma-8 \Gamma^\alpha_{\,\left[\sigma\alpha\right]} W^\sigma
\\
\nonumber
&+&4\Gamma^\alpha_{\,\left[ \lambda\alpha\right]}
\Gamma^\beta_{\,\left[\underline{\lambda}\beta\right]}- 2
\Gamma^\rho_{\,\left[\lambda \omega\right]}
\Gamma^\omega_{\,\left[\underline{\lambda} \rho\right]} -g_{\alpha\omega}
\Gamma^\alpha_{\,\left[\lambda\sigma\right]}
\Gamma^\omega_{\,\left[\underline{\lambda} \underline{\sigma}\right]} -
4\beta^2\Lambda \quad=0.
\end{eqnarray}
Finally varying in (\ref{13}) the gauging function $\beta$, one has
\begin{eqnarray}\label{24}
\beta R+k\beta_{;\alpha;\underline{\alpha}}= \beta (8
\Gamma^\alpha_{\,\left[\lambda \alpha\right]} w^\lambda-4
\Gamma^\alpha_{\,\left[\underline{\lambda} \alpha\right] ;\lambda}
-&4\Gamma^\alpha_{\,\left[\lambda \alpha\right]}
\Gamma^\beta_{\,\left[\underline{ \lambda}\beta\right]} &+2
\Gamma^\alpha_{\,\left[\lambda \rho\right]}
\Gamma^\rho_{\,\left[\underline{\lambda} \alpha\right]} +
g^{\mu\nu}g_{\alpha \omega}\Gamma^\alpha_{\,\left[\lambda \mu\right]}
\Gamma^\omega_{\,\left[\underline{\lambda}\nu\right]})
\\
\nonumber
+\beta (k-6)(w^\sigma w_\sigma-w^\sigma _{;\sigma}) +4 \beta^3\Lambda +
8\pi B \quad ,
\end{eqnarray}
with
\begin{equation}\label{25}
16\pi B=\frac{\delta L_{matter}}{\delta \beta} .
\end{equation}
{F}rom equations  (\ref{23}), and (\ref{24}) one obtains the
following relation
\begin{equation}\label{26}
(k-6)(\beta^2 W^\sigma)_{;\sigma}+8\pi(T-\beta B)+4(\beta^2  
\Gamma^\alpha_{\,\left[\underline{\sigma} \alpha\right]})  
_{\,;\sigma}
=0 \quad.
\end{equation}

Now, let us go back to the field equations (\ref{14}) -  
(\ref{19}).Contracting (\ref{17}) one has
\begin{equation}\label{27}
W^{\mu\nu}_{;\nu}=-3\beta^2W^\mu+ 2\beta^2
\Gamma^\nu_{\,\left[\underline{\mu} \nu\right]}  
-4\pi\Omega^{\,\left[\mu\nu\right]}_{\nu}
\quad .
\end{equation}
{F}rom (\ref{14}) and (\ref{27}) one has two interesting
conservation laws
\begin{equation}\label{28}
(k-6)(\beta^2W^\mu)_{;\mu}+8 \pi J^{\mu}_{\,;\mu}+ 4(\beta^2  
\Gamma^\nu_{\,\left[\underline{\mu} \nu\right]})_{\,;\mu}=0 \quad ,
\end{equation}
and
\begin{equation}\label{29}
3(\beta^2W^\mu)_{;\mu}+ 4\pi\Omega^{\,\left[\mu\nu\right]}_{\nu\,;\mu} -
2(\beta^2\Gamma^\nu _{\,\left[\underline{\mu} \nu\right]})_{;\mu}=  
0\quad.
\end{equation}

For the scalar $\int L_{matter}(-g)^{1/2} d^{4} x$ one can consider  
an infinitesimal transformation of coordinates. This leads to the  
following conservation law :
\begin{equation}\label{30}
T^{\sigma}_{\mu;\sigma}+ J^{\sigma}_{;\sigma} w_{\mu}  + J^{\sigma}  
W_{\mu\sigma}-B
\beta_{;\mu}- \Omega^{\,\left[\sigma \rho\right]} _{\lambda}
\Gamma^\lambda_{\,\left[\sigma \rho\right] ;\mu}+ (2
\Omega^{\,\left[\rho \sigma\right]} _{\lambda}
\Gamma^\lambda_{\,\left[\rho \mu\right]} - \Omega^{\,\left[\rho   
\lambda\right]}_{\mu}
\Gamma^\sigma_{\,\left[\rho \lambda\right]})_{;\sigma} =0 \quad .
\end{equation}
Considering for the same integral an infinitesimal gauge  
transformation,one  obtains the  relation
\begin{equation}\label{31}
J^{\mu}_{;\mu}= T-\beta B \quad.
\end{equation}
One readily sees that (\ref{31}) is compatible with
relations (\ref{26}), and (\ref{28}).

In order to get a non-vanishing magnetic current we will use
the well-known procedure (cf. e. g. \cite{Felsager},  
\cite{Landau})of building up  dual field equations.
With (\ref{19}) we can rewrite (\ref{14}) as follows
\begin{equation}\label{32}
\Phi^{\mu\nu}_{\; ; \nu}=(1/2)\beta^2 (k-6)W^\mu+
2\beta^2  \Gamma^\alpha_{\,\left[\underline{\mu}  
\alpha\right]}+4\pi J^\mu.
\end{equation}

Further, comparing between (\ref{14}) and (\ref{27}) we obtain
\begin{equation}\label{33}
2\Gamma^\alpha_{\,\left[\underline{\mu} \underline{\nu} \right];
\alpha;\nu} =
-(1/2) k \beta^2 W^\mu
- 4\pi(J^\mu+\Omega^{\,\left[\mu \nu\right]}_{\nu})\quad.
\end{equation}
We can  generalize expression (\ref{33}) writing :

\begin{equation} \label{34}
2\Gamma^\alpha_{\,\left[\underline{\mu}
\underline{\nu}\right];\alpha;\lambda}
= -(k/6) \beta^2(\delta^\nu_\lambda  W^\mu-\delta^\mu_\lambda    
W^\nu)- (4\pi/3) (J^\mu \delta^\nu_\lambda -J^\nu  
\delta^\mu_\lambda) -4\pi \Omega^{\,\left[\mu \nu\right]}_{\lambda}  
\quad.
\end{equation}

Making use of (\ref{19}), and (\ref{34}) we can write the
cyclic permutation
\begin{equation}\label{35}
\Phi_{\mu\nu;\lambda}+ \Phi_{\lambda\mu;\nu}
+ \Phi_{\nu\lambda;\mu} =
4\pi(\Omega_{\mu\left[\nu\lambda\right]}
+\Omega_{\lambda\left[\mu\nu\right]} +
\Omega_{\nu\left[\lambda\mu\right]})
 = 4\pi \Theta_{\mu\nu\lambda} \quad ,
\end{equation}
where $\Omega_{\lambda\left[\mu\nu\right]}
\equiv g_{\mu\alpha}g_{\nu\beta}
\Omega^{\,\left[\alpha \beta\right]}_{\lambda}$.

Now, let us define the dual field
\begin{equation}\label{36}
\widetilde{\Phi}^{\mu\nu} = -\frac{1}{2 (-g)^{1/2}}\;    
\varepsilon^{\mu\nu\alpha\beta}
 \Phi_{\alpha\beta} \quad ,
\end{equation}
and the current
\begin{equation}\label{37}
L^{\sigma}=- \frac{1}{6 (-g)^{1/2}}\; \varepsilon^{\sigma\mu\nu\lambda}
\Theta_{\mu \nu\lambda} \quad .
\end{equation}
where $\varepsilon^{\mu\nu\alpha\beta}$ stands for the
completely antisymmetric Levi-Civita symbol, and
$\varepsilon^{0123}=1$.( A book by O. Veblen \cite{Veblen},
containing useful information on Levi-Civita symbols,
must be mentioned here.)
For the dual of eq. (\ref{32}) we can now write:
\begin{equation}\label{38}
\widetilde{\Phi}^{\mu \nu}_{\; ; \nu}=4\pi L^{\mu} \quad .
\end{equation}
Finally, {F}rom (\ref{38}) one obtains
\begin{equation}\label{39}
L^{\mu}_{\; ;\mu}=0 \quad ,
\end{equation}
that may be interpreted as a conservation law for the
magnetic current density.

Equations (\ref{32}), and (\ref{38}) may be intepreted
as describing an electromagnetic field induced by the
electric current density $J^{\mu}$ and by the magnetic
current density $L^{\mu}$. These equations are covariant
under gauge transformations (\ref{7}), (\ref{9}).
It is worth noting that eq. (\ref{32}) is rather a
Proca equation then a Maxwell one. Turning to the
Einstein gauge $\beta=1$, considering a  torsionless geometry  
$\Gamma^{\lambda}_{\,\left[\mu\nu\right]}=0$,
(equation (\ref{35}), (\ref{37}), and (\ref{38}) vanish
identically in that case),  and taking $k=6$  we
obtain {F}rom (\ref{20}), and (\ref{32})  the  Einstein-Maxwell  
equations and {F}rom  eq. (\ref{30}) we obtain the energy  
conservation law in its conventional form.
\section{A  Possible  Theory}
In the previous section a general formalism was developed. In the  
present section we will specify it. Let us choose for the torsion  
tensor a  representation suitable for describing magnetism. We  
recall that in our case the Weyl connection vector $w_{\mu}$ acts as  
the electromagnetic potential vector. In order to describe the  
magnetic field and to  invoke a nonzero magnetic current density  we  
need  another vector. The torsion can
be broken in three irreducible parts
(cf. e.g.  \cite{Hammond} , \cite{Hayashi}): a trace part , a  
traceless one, and a totally antisymmetric part. Let us asumme that  
the first two parts vanish. In this case the totally antisymmetric  
torsion tensor
$\Gamma^{\lambda}_{\left[\mu\nu\right]}$ may be represented by a  
vector.If we
introduce the auxiliary torsion tensors:
\begin{equation}\label{40}
\Gamma_{\lambda\left[\mu \nu\right]}=g_{\sigma\lambda}
\Gamma^{\sigma}_{\left[\mu \nu\right]} ; \qquad
\Gamma^{\lambda\left[\mu \nu\right]}=g^{\alpha \mu}g^{\beta\nu}
\Gamma^{\lambda}_{\left[\alpha \beta\right]} \quad .
\end{equation}
we can express the torsion  by means of a vector $V^{\mu}$ as follows
\begin{equation}\label{41}
\Gamma_{\lambda\left[\mu\ nu\right]}=(-g)^{1/2} \;   
\varepsilon_{\lambda\mu \nu\sigma}V^{\sigma} ; \qquad
\Gamma^{\lambda\left[\mu \nu\right]}=-(-g)^{-1/2}\;  
\varepsilon^{\lambda\mu \nu\sigma}V_{\sigma}  \quad.
\end{equation}
This leads to
\begin{equation}\label{42}
\Gamma^{\nu}_{\left[\mu\nu\right]}=0  .
\end{equation}
{F}rom (\ref{41}) one can also derive the following simple formula:
\begin{equation}\label{43}
\Gamma^{\alpha\left[\mu\nu \right]}_{\;\; ;\alpha} =
\frac{\varepsilon^{\mu \nu\alpha\sigma}} {2(-g)^{1/2}}  
(V_{\alpha\,;\sigma}-
V_{\sigma \,;\alpha}).
\end{equation}
Making use of (\ref{43}) we can rewrite the following exact
expressions for the field tensors, defined by (\ref{19}) , and (\ref{36})
\begin{equation}\label{44}
\Phi^{\mu\nu}=(W^\mu_{;\,\underline\nu}- W^\nu_{;\,\underline\mu}) -
\frac{\varepsilon^{\mu\nu\alpha\sigma}} {(-g)^{1/2}}
(V_{\alpha\,;\,\sigma}- V_{\sigma\,;\,\alpha}) \quad ,
\end {equation}
and
\begin{equation}\label{45}
\widetilde\Phi^{\mu\nu}= -2(V^\mu_{;\,\underline\nu}    
-V^\nu_{;\,\underline\mu}) -
\frac{\varepsilon^{\mu\nu \alpha\sigma}}
{2(-g)^{1/2}}(W_{\alpha\,;\,\sigma} -
W_{\sigma\,;\,\alpha}) \quad.
\end {equation}
Inserting these into equations (\ref{32}) and (\ref{38}),
and making use of  (\ref{42}), we obtain
\begin{equation} \label{46}
W^\mu_{;\underline{\nu} ;\nu} -W^\nu_{;\underline{\mu};\nu} -
\frac{\varepsilon^{\mu\nu\alpha\sigma}}
{(-g)^{1/2}}(V_{\alpha , \sigma , \nu}
-V_{\sigma , \alpha , \nu}) =(1/2)(k-6)\beta^2 W^\mu+ 4 \pi J^\mu \quad,
\end {equation}
  and
\begin{equation}\label{47}
-2(V^\mu_{;\,\underline{\nu} ; \,\nu}  -V^\nu_{;\,\underline{\mu};  
\nu}) -
\frac{\varepsilon^{\mu\nu\alpha\sigma}} {2(-g)^{1/2}}
(W_{\alpha , \sigma , \nu} -W_{\sigma , \alpha ,\nu})= 4\pi L^\mu \quad.
\end {equation}
One can readily see that in (\ref{46}), and in (\ref{47}) the  
termswith the
Levi-Civita symbols vanish identically, so that one is left with
the following
field equations
\begin{equation}\label{48}
\Phi^{\mu\nu}_{\, ; \nu} =  W^\mu_{;\underline{\nu}; \nu}  
-W^\nu_{;\underline{\mu}; \nu}
= (1/2)(k-6)\beta^2  W^\mu+4 \pi J^\mu \quad,
\end {equation}
and

\begin{equation}\label{49}
\widetilde\Phi^{\mu\nu}_{\, ; \nu} =(V^{\mu}_{;\,\underline{\nu}  
;\, \nu} - V^{\nu}_{;\, \underline{\mu}; \nu})
 =-2 \pi L^{\mu}
\quad.
\end{equation}
{F}rom (\ref{48}) one sees that
\begin{equation}\label{50}
\Phi^{\mu \nu}_{\, ;\,\nu} =W^{\mu \nu}_{\,; \,\nu}  .
\end {equation}
Thus, comparing (\ref{48}) with (\ref{27}), and taking into
account (\ref{42}) , and (\ref{50}) one obtains
\begin{equation}\label{51}
\Omega^{\,\left[\mu\nu\right]}_{\nu} =J^{\mu}+(1/8\pi) \beta^2 k W^{\mu}.
\end {equation}

Now let us consider the
tensor $\Omega^{\,\left[\mu\nu\right]}_{\lambda}$
as defined by (\ref{18}).  It has to satisfy condition (\ref{51}),
and its structure must be in accordance with (\ref{41}).
It is convenient to introduce auxiliary tensors
\begin{equation}\label{52}
\Omega^{\lambda\left[\mu \nu\right]}
= g^{\lambda\sigma} \Omega^{\,\left[\mu \nu\right]}_{\sigma}; \qquad
 \Omega_{\lambda\left[\mu \nu\right]} =
g_{\alpha\mu} g_{\beta\nu} \Omega^{\,\left[\alpha  
\beta\right]}_{\lambda}   \quad.
\end{equation}
An appropriate choice is
\begin{equation}\label{53}
\Omega_{\lambda\left[\mu \nu\right]}=(1/3)(g_{\nu \lambda}J_{\mu}-
g_{\mu \lambda}J_{\nu})+ \frac{k \beta^2}{24\pi}(g_{\nu \lambda} W_{\mu}-
g_{\mu \lambda}W_{\nu}) +(-g)^{1/2}\varepsilon_{\lambda  
\mu\nu\sigma} l^{\sigma}
\quad,
\end{equation}
and
\begin{equation}\label{54}
\Omega^{\lambda\left[\mu \nu\right]}=(1/3)(g^{\nu \lambda} J^{\mu}  
-g^{\mu \lambda}J^{\nu})+\frac{k \beta^2}{24 \pi}(g^{\nu  
\lambda}W^{\mu} -
g^{\mu\lambda}W^{\nu})
- (-g)^{-1/2}\varepsilon^{\lambda \mu\nu\sigma} l_{\sigma} \quad;
\end{equation}
 with $l_{\mu}$ being a gauge invariant vector.

Making use of (\ref{40}) - (\ref{42}) one finds that
equation (\ref{20}) now takes on the form
\begin{eqnarray}\label{55}
G^{\mu\nu}= -(8\pi/{\beta^2}) T^{\mu\nu}-(8\pi/{\beta^2})  
(\widetilde{M}^{\mu\nu}-\overline {M}^{\mu\nu}) +
(2/\beta)(g^{\mu \nu} \beta_{\, ;\underline{\alpha} \,;\alpha} -
\beta_{\, ;\underline{\mu}\, ; \underline{\nu}})
\\
\nonumber
+(1/\beta^2) (4\beta_{; \underline{\mu}}
\beta_{; \underline{\nu}} -g^{\mu\nu}
\beta_{; \underline{\alpha}}
\beta_{;\alpha})
+ (k-6)(W^\mu  W^\nu -\frac{1}{2} g^{\mu\nu} W^\sigma W_\sigma)  
-g^{\mu\nu} V^\sigma V_\sigma-2V^\mu V^\nu .
\end{eqnarray}
The equations describing the electromagnetic
fields are given by  (\ref{48}),
(\ref{49}), and for the magnetic current density
vector $L^{\mu}$ one has by
(\ref{35}), (\ref{37}), and (\ref{53})
\begin{equation}\label{56}
L^\sigma =3l^\sigma
\end{equation}
{F}rom (\ref{48}), (\ref{49}) one can  also derive a current
conservation law
\begin{equation}\label{57}
(k-6)(\beta^2 W^{\mu})_{\, ;\mu} + 8\pi J^{\mu}_{\, ; \mu}=0 ,
\end{equation}
as well  the conservation law for the magnetic currrent  (\ref{39}).

It is remarkable that in (\ref{48}) , (\ref{49})  the electric  
currentdensity vector $J^\mu$ appears as a creator of the $W$-part  
of the field,
while the magnetic current density vector $L^\mu$ creates the $V$-part.
\section{The Einstein Gauge}
The torsionless Weyl-Dirac theory with $k=6$ takes on the form of  
the Einstein-Maxwell theory if one choose the Einstein gauge  
$\beta=1$. (cf.\cite{Dirac1973},\cite{Rosen1982},  
\cite{Isr.Rosen1983}). Thus the
Eistein gauge may help to  understanding the theory.

In this section we consider the theory, possessing torsion, and  
allowing arbitrary values for $k$, which was developed in the  
previous section.
Turning to the Einstein gauge, we set
\begin{equation}\label{58}
\beta=1 .
\end{equation}
 By condition (\ref{58}) we also have $w^{\mu}$ instead $W^{\mu}$,  
andequation (\ref{55}) now takes on the form
\begin{eqnarray}\label{60}
G^{\mu\nu}=- 8\pi T^{\mu\nu} -8\pi (\widetilde{M}^{\mu\nu}  
-\overline{M}^{\mu\nu})- (k-6)(w^{\mu} w^{\nu}- (1/2)g^{\mu\nu}  
w^{\sigma} w_{\sigma})
\\
\nonumber
-2V^{\mu} V^{\nu}- g^{\mu\nu} V^{\sigma} V_{\sigma} \quad .
\end {eqnarray}
The energy conservation  law  can  be obtained from (\ref{30}) by
complicated and very lengthy calculations.  Alternatively one can,
making use of the Bianchi identities, and taking the divergence,
derive  this law from equation (\ref{60}).

Let us choose the second way. We have  $G^{\nu}_{\mu ; \nu}=0$ , so  
that(\ref{60}) leads to
\begin{equation}\label{61}
8\pi ( T^{\nu}_{\mu ;\,\nu}+ \widetilde{M}^{\nu}_{\mu ; \, \nu} -
\overline{M}^{\nu}_{\mu ;\, \nu})+ (k-6)\left(w_{\mu} w^{\nu}-(1/2)  
\delta^{\nu}_{\mu} w_{\sigma} w^{\sigma}\right)_{;\, \nu}
+ 2(V_{\mu} V^{\nu}) _{;\,\nu}+ (V_{\sigma} V^{\sigma})_{;\,\mu} =0 .
\end{equation}
Making in (\ref{61}) use of definitions (\ref{21}), and (\ref{22}),  
and of equations (\ref{35}), (\ref{48}) , (\ref{50}), (\ref{53}),  
(\ref{56}) one obtains
\begin{eqnarray}\label{62}
8\pi( T^{\nu}_{\mu \, ; \, \nu}+\Psi_{\mu\sigma} J^{\sigma}) +4\pi   
\sqrt{-g}\,\,
 \varepsilon_{\alpha \beta\mu \sigma} W^{\alpha\beta} L^{\sigma}+ (k-6)
(\Psi_{\mu\sigma}+ W_{\mu\sigma})w^{\sigma}
\\
\nonumber
+(k-6) w_{\mu} w^{\nu}_{; \,\nu}
+   2V^{\nu} (V_{\mu \, ; \, \nu}+V_{\nu \, ; \, \mu}) +
2V_{\mu} V^{\nu}_{\, ; \,\nu} =0 \quad.
\end {eqnarray}
For a moment let us go back to (\ref{48}). With
a new parameter $\kappa^2 \equiv (1/2)(6-k)$ we can rewrite this
equation as
\begin{equation}\label{63}
\Phi^{\mu\nu}_{\, ; \nu} =  w^\mu_{;\underline{\nu};\nu}-   
w^\nu_{;\underline{\mu};\nu}
= -\kappa^{2} w^\mu+4 \pi J^\mu \quad.
\end {equation}
In absent of electric currents in a certain region we
obtain {F}rom (\ref{57}), and (\ref{58})
\begin{equation}\label{64}
w^{\nu}_{\, ;\,\nu}=0 \quad,
\end {equation}
so that equation (\ref{63}) may be rewritten in the following form
\begin{equation}\label{65}
w^{\mu}_{ ;\,\underline{\nu};\, \nu} +w^\nu R_{\nu}^{\mu}
+ \kappa^2w^\mu=0   \quad;
\end {equation}
with $R_{\nu}^{\mu}$ being the Ricci tensor, formed from the
usual Christoffel symbols. If the curvature in the current-free
region  is negligible, one obtains  the Proca \cite{Proca1936}
equation for the vector field $w^{\mu}$
\begin{equation}\label{66}
w^{\mu}_{ ;\,\underline{\nu};\, \nu}+ \kappa^2  w^\mu = 0;
\end {equation}

{F}rom the quantum mechanical standpoint this equation describes a
particle having spin $1$ and mass that in  conventional units is
given by
\begin{equation}\label{67}
m=(\hbar/c)\kappa= (\hbar/c)\sqrt{\frac{6-k}{2}} ,
\end {equation}
thus, for $k<6$ one obtains massive field particles, photons.

In the special case when, $V^{\mu}=0$, and $k=6$, one obtains   
{F}rom (\ref{49}) $L^{\mu}=0$, so that equations (\ref{60}), and  
(\ref{63})
turn into the   equations of  the Einstein-Maxwell theory,
while  (\ref{62}) becomes the  usual energy conservation law.
\section{Discussion}
The Weyl geometry \cite{Weyl1919} is doubtless the most aesthetic  
generalization of the Riemannian geometry, the last being the  
framework of general relativity.
Dirac \cite{Dirac1973} modified the Weyl theory. In order to build  
upan action integral, which is coordinate invariant, and  gauge  
invariant,and which agrees with the general relativity theory, Dirac  
 introduced a scalar gauging function,  $\beta$. The modified   
Weyl-Dirac theoryoffers a complete basis for deriving  gravitation,  
and
electromagnetism {F}rom geometry (cf. e.g. \cite{Isr.Rosen1983},  
\cite{Israelit1989}).

Generalizing the Weyl-Dirac framework \cite{Weyl1919},  
\cite{Dirac1973}, \cite{Rosen1982}, we have developed in the present  
work a geometrically
based general relativistic theory, possessing intrinsic electric  
andmagnetic currents and admitting massive photons. Torsion is  
includedinto the geometry, so that  the linear connection  
(cf.(\ref{1}))
is made now of three  parts,  the metric  (Christoffel) term, the
Weylian non-metricity, and contorsion. Following Dirac   
\cite{Dirac1973} we also make use of the gauging function.

The general procedure is given in section  3. Varying in the  
actionthe metric tensor $g_{\mu\nu}$  , the Weyl connection vector  
$w_{\mu}$ ,the torsion tensor $\Gamma^{\lambda}_{\;  
\left[\mu\nu\right]}$,
and the gauge function $\beta$ one obtains the equations,  as well  
as conservation laws. Making use of these equations one can  
introduce
such a field tensor (cf. (\ref{19}) )  that its dual (\ref{36})  
defines a nonzero magnetic current density vector. The  
eletromagnetic field is
related to the electric current by a Proca eq.  (cf. (\ref{32}) )
instead of a Maxwell one, so that massive photons are expected.
However one can choose the value of an arbitrary parameter to get
massless photons and Maxwell equations. The whole theory is gauge
covariant in the Weyl sense.

In section 4 we specify the torsion tensor, assuming that it has a
totally antisymmetric structure. Making this assumption enables to
represent the torsion as a dual of a vector field.  This  
representationresults in a simple form of the field equations. One  
obtains also charge conservation laws  (\ref{39}) ,  (\ref{57}).  
{F}rom the field equation (\ref{48}) one sees that the  
electromagnetic potential
vector   $W_{\mu}$ has two sources, the electric current density
vector $J_{\mu}$ , and a Proca term, while according to (\ref{49})
the magnetic vector $V_{\mu}$ is created only by magnetic
currents $L_{\mu}$ . The two field vectors are coupled by the
energy conservation law  and by the equation for the gravitation  
field (\ref{55}) (cf. definitions  (\ref{21}), (\ref{22})).

Assuming $\beta=1$ one turns to the Einstein gauge, in which the
field, and its dual are given by (\ref{63}), and  (\ref{49})
respectively . The coupling of these two fields and their currents   
is manifested in the energy conservation law (\ref{62}).  So one  
sees
that the magnetic current vector $L^{\mu}$ is coupled to the  
electricfield $W^{\mu\nu}$, while the electric current  $J^{\mu}$ is  
coupledto the whole field. The mass term (with the factor $(k-6)$)   
is alsoconnected to both fields. One also  obtains an equation  
(\ref{66})
that may be treated as describing   massive field particles,  
photonsin  current-free regions. The photon mass is fixed by an  
arbitrary
parameter $\kappa$  that may be chosen to give any extremely small  
value for  $m_{\gamma}$. Choosing a vanishing photon mass we
obtain {F}rom (\ref{63}) the classical Maxwell equation,  
possessingthe usual Maxwell gauge invariance.

A geometric based theory is proposed. It is shown that electric,
and intrinsic magnetic currents, as well massive photons  coexist
within this framework.  If one drops the torsion , and sets $k=6$
one has the Weyl-Dirac theory, which in the Einstein gauge takes
on the form of Einstein-Maxwell.

\section*{Acknowledgments}
{The author takes this opportunity to express his cordial thanks to  
Professor HEINZ DEHNEN for very interesting discussions.}


\begin{thebibliography}{99}
\bibitem{Dirac1931}
P.~A.~M.~Dirac,~ Proc.~Roy.~Soc. {\bf A133}, 60, (1931).
\bibitem{Dirac1948}
P.~A.~M.~Dirac, ~Phys.~Rev. {\bf74}, 817, (1948).
\bibitem{Blagojevic}
M.~Blagojevi\'{c} , and P.~Senjanovi\'{c} , Physics Reports,  
{\bf157}, 234, (1988).
\bibitem{Hammond}
R.~T.~Hammond, Il~ Nuovo~ Cimento {\bf 108~B}, 725, (1993).
\bibitem{Goldhaber}
A.~S.~Golhaber, and~M.~M.~Nieto , Rev.~Mod.~Phys. {\bf43}, 277, (1971).
\bibitem{Ignatiev}
A.~Yu.~Ignatiev, and ~G.~C.~Joshi, Phys.~Rev.~D. {\bf53}, 984, (1996).
\bibitem{Weyl1919}
H.~Weyl,~ Ann.~Phys.~(Leipzig)~{\bf 59}, 101, (1919).
\bibitem{Dirac1973}
P.~A.~M.~Dirac, ~Proc.~Roy.~Soc. {\bf A333}, 403, (1973).
\bibitem{Schouten}
J.~Schouten, {\em Ricci Calculus},(Springer Verlag, Berlin, 1954).
\bibitem{Rosen1982}
N.~Rosen,~Found.~Phys. {\bf12}, 213, (1982).
\bibitem{Isr.Rosen1992}
M.~Israelit and~ N.~Rosen,~Found.~Phys. {\bf22}, 555, (1992).
\bibitem{Isr.Rosen1994}
M.~Israelit and~N.~Rosen, Found.~Phys. {\bf24}, 901, (1994).
\bibitem{Felsager}
B.~Felsager, {\em Geometry, Particles and Fields}, (Odense  
University Press 1981).
\bibitem{Landau}
L.~D.~Landau and E.~M.~Lifshitz,  {\em The Classical Theory of Fields},
(Pergamon, Oxford, 1975).
\bibitem{Proca1936}
A.~L.~Proca, J.~Phys.~Rad. {\bf 7}, 347, (1936)
\bibitem{Veblen}
O.~Veblen,  {\em Invariants of Quadratic Differential Forms } ,
( University Press, Cambridge, 1962).
\bibitem{Isr.Rosen1983}
M.~Israelit and~N.~Rosen, Found.~Phys. {\bf13}, 1023, (1983).
\bibitem{Israelit1989}
M.~Israelit, Found.~Phys.  {\bf19}, 35, (1989)
\bibitem{Hayashi}
K.~Hayashi and T.~ Shirafuji, Prog.~Theor.~Phys., {\bf64}, 866, (1980).
\end{thebibliography}
\end{document}